\documentclass[12pt,a4paper]{article}
\usepackage{graphicx}
\usepackage{subfigure}
\usepackage{amsmath}
\usepackage{amsthm}
\usepackage{amssymb}
\usepackage{url}
\newtheorem{thm}{Theorem}

\newtheorem{lem}[thm]{Lemma}
\newtheorem{prop}[thm]{Proposition}
\theoremstyle{definition}
\newtheorem{exmp}{Example}
\newtheorem{defin}{Definition}
\newcommand{\maxsize}[1]{\mathrm{maxsize}(#1)}

\graphicspath{{experimenten/}}

\title{A tight upper bound \\ on the number of candidate patterns\footnote{A preliminary report on this work was presented at the
2001 IEEE International Conference on Data Mining~\cite{bounds}.}}
\author{{\sl Floris Geerts}, {\sl Bart Goethals}, {\sl Jan Van den Bussche}\\
University of Limburg, Belgium}
\date{\ }

\begin{document}
\begin{titlepage}
\maketitle
\thispagestyle{empty}

\begin{abstract}
In the context of mining for frequent patterns using the standard
levelwise algorithm, the following question arises: given the
current level and the current set of frequent patterns, what is
the maximal number of candidate patterns that can be generated on
the next level? We answer this question by providing a tight upper
bound, derived from a combinatorial result from the sixties by
Kruskal and Katona. Our result is useful to reduce the number of
database scans.
\end{abstract}

\end{titlepage}

\section{Introduction} \label{intro}

The frequent pattern mining problem \cite{ais} is by now well
known. We are given a set of items $\cal I$ and a database $\cal
D$ of subsets of $\cal I$ called transactions. A \emph{pattern} is
some set of items; its \emph{support} in $\cal D$ is defined as
the number of transactions in $\cal D$ that contain the pattern;
and a pattern is called \emph{frequent} in $\cal D$ if its support
exceeds a given minimal support threshold. The goal is now to find
all frequent patterns in $\cal D$.

The search space of this problem, all subsets of $\cal I$, is
clearly huge. Instead of generating and counting the supports of
all these patterns at once, several solutions have been proposed
to perform a more directed search through all patterns.  However,
this directed search enforces several scans through the database,
which brings up another great cost, because these databases tend
to be very large, and hence they do not fit into main memory.

The standard Apriori algorithm \cite{kddboek_chap12} for solving
this problem is based on its monotonicity property, that all
subsets of a frequent pattern must be frequent. A pattern is thus
considered potentially frequent, also called a \emph{candidate}
pattern, if its support is yet unknown, but all of its subsets are
already known to be frequent. In every step of the algorithm, all
candidate patterns are generated and their supports are then
counted by performing a complete scan of the transaction database.
This is repeated until no new candidate patterns can be generated.
Hence, the number of scans through the database equals the maximal
size of a candidate pattern. Several improvements on the Apriori
algorithm try to reduce the number of scans through the database
by estimating the number of candidate patterns that can still be
generated.

At the heart of all these techniques lies the following purely
combinatorial problem, that must be solved first before we can
seriously start applying them:
\emph{given the current set of frequent patterns at a certain pass
of the algorithm, what is the maximal number of candidate patterns
that can be generated in the passes yet to come?}

Our contribution is to solve this problem by providing a hard and
tight combinatorial upper bound. By computing our upper bound
after every pass of the algorithm, we have at all times a
watertight guarantee on the size of what is still to come, on
which we can then base various optimization decisions, depending
on the specific algorithm that is used.

In the next Section, we will discuss existing techniques to reduce
the number of database scans, and point out the dangers of using
existing heuristics for this purpose.  Using our upper bound,
these techniques can be made watertight. In Section~\ref{basicbounds}, we derive
our upper bound, using a combinatorial result from the sixties by
Kruskal and Katona.  In Section~\ref{sterbounds}, we show how to get even more
out of this upper bound by applying it recursively.  We will then
generalize the given upper bounds such that they can be applied by
a wider range of algorithms in Section~\ref{generalbounds}.
In Section~\ref{implement}, we discuss
several issues concerning the implementation of the given upper
bounds on top of Apriori-like algorithms. In Section~\ref{experiment}, we give
experimental results, showing the effectiveness of our result in
estimating, far ahead, how much will still be generated in the
future. Finally, we conclude the paper in Section~\ref{conclusion}.

\section{Related Work} \label{related}

Nearly all frequent pattern mining algorithms developed after the
proposal of the Apriori algorithm, rely on its levelwise candidate
generation and pruning strategy.  Most of them differ in how they
generate and count candidate patterns.

One of the first optimizations was the DHP algorithm proposed by
Park et al.\ \cite{dhp}. This algorithm uses a hashing scheme to
collect upper bounds on the frequencies of the candidate patterns
for the following pass.  Patterns of which it is already known
that they will turn up infrequent can then be eliminated from
further consideration. The effectiveness of this technique only
showed for the first few passes. Since our upper bound can be used
to eliminate passes at the end, both techniques can be combined in
the same algorithm.

Other strategies, discussed next, try to reduce the number of passes.  However,
such a reduction of passes often causes an increase in the number
of candidate patterns that need to be explored during a single
pass. This tradeoff between the reduction of passes and the number
of candidate patterns is important since the time needed to
process a transaction is dependent on the number of candidates
that are covered in that transaction, which might blow up
exponentially. Our upper bound can be used to predict whether or not this blowup will occur.

The Partition algorithm, proposed by Savasere et al.\
\cite{partition}, reduces the number of database passes to two.
Towards this end, the database is partitioned into parts small
enough to be handled in main memory.  The partitions are then
considered one at a time and all frequent patterns for that
partition are generated using an Apriori-like algorithm. At the
end of the first pass, all these patterns are merged to generate a
set of all potential frequent patterns, which can then be counted
over the complete database.  Although this method performs only
two database passes, its performance is heavily dependent on the
distribution of the data, and could generate much too many candidates.

The sampling algorithm proposed by Toivonen \cite{sampling}
performs at most two scans through the database by picking a
random sample from the database, then finding all frequent
patterns that probably hold in the whole database, and then
verifying the results with the rest of the database. In the cases
where the sampling method does not produce all frequent patterns,
the missing patterns can be found by generating all remaining
potentially frequent patterns and verifying their frequencies
during a second pass through the database.  The probability of
such a failure can be kept small by decreasing the minimal support
threshold.  However, for a reasonably small probability of failure,
the threshold must be drastically decreased,
which can again cause a combinatorial explosion of
the number of candidate patterns.

The DIC algorithm, proposed by Brin et al.\ \cite{dic}, tries to
reduce the number of passes over the database by dividing the
database into intervals of a specific size.  First, all candidate
patterns of size $1$ are generated.  The frequencies of the
candidate sets are then counted over the first interval of the
database.  Based on these frequencies, candidate patterns of size
$2$ are generated and are counted over the next interval together
with the patterns of size $1$.  In general, after every interval
$k$, candidate patterns of size $k+1$ are generated and counted.
The algorithm stops if no more candidates can be generated. Again,
this technique can be combined with our technique in the same
algorithm.

Another type of algorithms generate frequent patterns using a
depth-first search~\cite{eclat, depthproject, treeproject,
fpgrowth}. Generating patterns in a depth-first manner implies
that the monotonicity property cannot be exploited anymore. Hence,
a lot more patterns will be generated and need to be counted,
compared to the breadth-first algorithms. The FPgrowth algorithm
from Han et al.\ solves this problem by loading a compressed form
of the database in main memory using the proposed FPtree.  This
memory-resident FPtree benefits from a very fast counting
mechanism of all generated patterns.\footnote{Note that the
patterns in the FPtree are represented in the so called header
tables.} Obviously, it is not always possible to load the
compressed form of the database into main memory.

Other strategies try to push certain constraints into the
candidate pattern generation as deeply as possible to reduce the
number of candidate patterns that must be
generated~\cite{interactive, lnhp_cfq, nlhp_car, sva_car}. Still
others try to find only the set of \emph{maximal} frequent
patterns, i.e. those frequent patterns that have no superset which
is also frequent \cite{maxminer, eclat, pincer}.  Of course, these
techniques do not give us all frequencies of all frequent patterns
as required by the general pattern mining problem we consider in
this paper.

The first heuristic specifically proposed to estimate the number
of candidate patterns that can still be generated was used in
the AprioriHybrid algorithm~\cite{apriori,apriorihybrid}. This
algorithm uses Apriori in the initial iterations and switches to
AprioriTid if it expects it to run faster. This AprioriTid
algorithm does not use the database at all for counting the
support of candidate patterns. Rather, an encoding of the
candidate patterns used in the previous iteration is employed for this
purpose.  The AprioriHybrid algorithm switches to AprioriTid when
it expects this encoding of the candidate patterns to be small
enough to fit in main memory. The size of the encoding grows with
the number of candidate patterns. Therefore, it calculates the
size the encoding would have in the current iteration. If this size is
small enough and there were fewer candidate patterns in the
current iteration than the previous iteration, the heuristic decides to
switch to AprioriTid.

This heuristic (like all heuristics) is not waterproof, however.
Take, for example, two disjoint datasets. The first dataset
consists of all subsets of a frequent pattern of size $20$.  The
second dataset consists of all subsets of $1\,000$ disjoint
frequent patterns of size $5$.  If we merge these two datasets, we
get $\binom{20}{3} + 1\,000 \binom{5}{3} = 11\,140$ patterns of
size $3$ and $\binom{20}{4} + 1\,000 \binom{5}{4} = 9\,845$
patterns of size $4$.  If we have enough memory to store the
encoding for all these patterns, then the heuristic decides to
switch to AprioriTid. This decision is premature, however, because
the number of new patterns in each pass will start growing
exponentially afterwards.

Also, current state-of-the-art algorithms for frequent itemset
mining, such as Opportunistic Project~\cite{han:op} and
DCI~\cite{dci} use several techniques within the same algorithm
and switch between these techniques using several simple, but not
waterproof heuristics.

Another improvement of the Apriori algorithm, which is part of the
folklore, tries to combine as many iterations as possible in the
end, when only few candidate patterns can still be generated. The
potential of such a combination technique was realized early on
\cite{apriori,kddboek_chap12}, but the modalities under which it
can be applied were never further examined.  Our work does exactly
that.

\section{The basic upper bounds} \label{basicbounds}

In all that follows, $L$ is some family of patterns of size $k$.

\begin{defin}
A \emph{candidate pattern} for $L$ is a pattern (of size larger
than $k$) of which all $k$-subsets are in $L$. For a given $p
> 0$, we denote the set of all size-$k+p$ candidate patterns
for $L$ by $C_{k+p}(L)$.
\end{defin}

For any $p \geq 1$, we will provide an upper bound on
$|C_{k+p}(L)|$ in terms of $|L|$.  The following lemma is central
to our approach. (A simple proof was given by
Katona~\cite{katona}.)

\begin{lem}
Given $n$ and $k$, there exists a unique representation $$ n=\binom{m_k}{k} + \binom{m_{k-1}}{k-1} + \cdots + \binom{m_r}{r}, $$
with $r \geq 1$, $m_k>m_{k-1}>\ldots>m_r$, and $m_i\geq i$ for
$i=r,r+1,\ldots,k$.
\end{lem}
This representation is called the $k$-\textit{canonical
representation of} $n$ and can be computed as follows: The integer
$m_k$ satisfies $\binom{m_k}{k} \leq n < \binom{m_k+1}{k}$, the
integer $m_{k-1}$ satisfies $\binom{m_{k-1}}{k-1} \leq
n-\binom{m_k}{k} < \binom{m_{k-1}+1}{k-1}$, and so on, until $n -
\binom{m_k}{k} - \binom{m_{k-1}}{k-1} - \cdots - \binom{m_r}{r}$
is zero.

We now establish:

\begin{thm} \label{kk}
If $$ |L|= \binom{m_{k}}{k} + \binom{m_{k-1}}{k-1} +\cdots +
\binom{m_r}{r} $$ in $k$-canonical representation, then
\begin{equation*}
|C_{k+p}(L)| \leq \binom{m_{k}}{k+p} + \binom{m_{k-1}}{k-1+p}
+\cdots + \binom{m_{s+1}}{s+p+1},
\end{equation*}
where $s$ is
the smallest integer such that $m_{s}<s+p$. If no such integer
exists, we set $s=r-1$.
\end{thm}
\begin{proof}
Suppose, for the sake of contradiction, that
$$
|C_{k+p}(L)| \geq \binom{m_{k}}{k+p} +
\binom{m_{k-1}}{k-1+p}
 +\cdots +
\binom{m_{s+1}}{s+p+1} + \binom{s+p}{s+p}. $$ Note that this is in
$k+p$-canonical representation. A theorem by Kruskal and Katona
\cite{frankl,katona,kruskal} says that $$ |L| \geq
\binom{m_{k}}{k} + \binom{m_{k-1}}{k-1} +\cdots +
\binom{m_{s+1}}{s+1}+ \binom{s+p}{s}. $$ But this is impossible,
because
\begin{align*}
|L| & =  \binom{m_{k}}{k} + \binom{m_{k-1}}{k-1} +\cdots +
        \binom{m_{s+1}}{s+1}
         +\binom{m_s}{s}+ \cdots + \binom{m_r}{r}\\
    &
    \leq  \binom{m_{k}}{k} + \binom{m_{k-1}}{k-1} +\cdots
+\binom{m_{s+1}}{s+1}
      + \sum_{1\leq i\leq s}\binom{i+p-1}{i}\\
    & <  \binom{m_{k}}{k} + \binom{m_{k-1}}{k-1} +\cdots
+\binom{m_{s+1}}{s+1}
      + \sum_{0\leq i\leq s}\binom{i+p-1}{i}\\
    & =  \binom{m_{k}}{k} + \binom{m_{k-1}}{k-1} +\cdots
        + \binom{m_{s+1}}{s+1} + \binom{s+p}{s}.
\end{align*}
The first inequality follows from the observation that $m_s\leq s+p-1$
implies $m_i\leq i+p-1$ for all $i=s,s-1,\ldots,r$.
The last equality follows from a well-known binomial identity.
\end{proof}

\paragraph{Notation} We will refer to
the upper bound provided by the above theorem as ${\it
KK}_k^{k+p}(|L|)$ (for Kruskal-Katona).  The subscript $k$, the
level at which we are predicting, is important, as the only
parameter is the cardinality $|L|$ of $L$, not $L$ itself.  The
superscript $k+p$ denotes the level we are predicting.

\begin{prop}[Tightness]
The upper bound provided by Theorem~\ref{kk} is \emph{tight:}
for any given $n$ and $k$ there always exists an $L$ with $|L| =
n$ such that for any given $p$, $|C_{k+p}(L)| = {\it
KK}_k^{k+p}(|L|)$.
\end{prop}
\begin{proof}
Let us write a finite set of natural numbers as a string of
natural numbers by writing its members in decreasing order.  We
can then compare two such sets by comparing their strings in
lexicographic order. The resulting order on the sets is known as
the \emph{colexicographic} (or \emph{colex}) order.  An
intuitive proof of the Kruskal-Katona theorem, based on this colex
order, was given by Bollob\'as \cite{bollo_comb}. Let
$$\binom{m_{k}}{k} + \binom{m_{k-1}}{k-1} +\cdots +
\binom{m_{r}}{r}$$ be the $k$-canonical representation of $n$.
Then, Bollob\'as has shown that all $k-p$-subsets of the first $n$
$k$-sets of natural numbers in colex order, are exactly the first
$$\binom{m_{k}}{k-p} + \binom{m_{k-1}}{k-1-p} +\cdots +
\binom{m_{s}}{r-s}$$ $k-p$-sets of natural numbers in colex order,
with $s$ the smallest integer such that $s>p$. Using the same
reasoning as above, we can conclude that all $k+p$-supersets of
the first $n$ $k$-sets of natural numbers in colex order are
exactly the first $KK_k^{k+p}(n)$ $k+p$-sets of natural numbers in
colex order.
\end{proof}
Analogous tightness properties hold for all upper bounds we will
present in this paper, but we will no longer explicitly state
this.

\begin{exmp}
Let $L$ be the set of $13$ patterns of size $3$:
$$\begin{array}{l} \{\{3,2,1\},\{4,2,1\},\{4,3,1\},\{4,3,2\}, \\
\{5,2,1\},\{5,3,1\},\{5,3,2\},\{5,4,1\},\{5,4,2\},\{5,4,3\}, \\
\{6,2,1\},\{6,3,1\},\{6,3,2\}\}. \end{array}$$ The $3$-canonical
representation of $13$ is $\binom{5}{3} + \binom{3}{2}$ and hence
the maximum number of candidate patterns of size $4$ is ${\it
KK}_3^{4}(13) = \binom{5}{4} + \binom{3}{3} = 6$ and the maximum
number of candidate patterns of size $5$ is ${\it KK}_3^{5}(13) =
\binom{5}{5} = 1$. This is tight indeed, because
\begin{multline*}
C_{4}(L) = \{\{4,3,2,1\},\{5,3,2,1\},\{5,4,2,1\},\\ \{5,4,3,1\},\{5,4,3,2\},\{6,3,2,1\}\}
\end{multline*}
and $$C_5(L) = \{\{5,4,3,2,1\}\}.$$
\end{exmp}

\paragraph{Estimating the number of levels}
The $k$-canonical representation of $|L|$ also yields an upper bound on the
maximal size of a candidate pattern, denoted by $\maxsize{L}$.  Recall that
this size equals the number of iterations the standard Apriori algorithm will
perform.  Indeed, since $|L| < \binom{m_k + 1}{k}$, there cannot be a
candidate pattern of size $m_k + 1$ or higher, so:

\begin{prop} \label{muprop} If $\binom{m_k}{k}$ is the first term in the $k$-canonical
representation of $|L|$, then $\maxsize{L} \leq m_k$.
\end{prop}
We denote this number $m_k$ by $\mu_k(|L|)$. From the form of
${\it KK}_k^{k+p}$ as given by Theorem~\ref{kk}, it is immediate
that $\mu$ also tells us the last level before which ${\it KK}$
becomes zero. Formally:
\begin{prop} \label{boils}
\begin{equation*}
\mu_k(|L|) = k + \min \{ p \mid {\it KK}_k^{k+p}(|L|) = 0 \} -1.
\end{equation*}
\end{prop}
\paragraph{Estimating all levels}
As a result of the above, we can also bound, at any given level $k$, the
\emph{total} number of candidate patterns that can be generated, as follows:
\begin{prop} \label{kktotal}
The total number of candidate patterns that can be generated from
a set $L$ of $k$-patterns is at most
\begin{equation*}
\mathit{KK}_k^{\rm total}(|L|) := \sum_{p \geq 1}
\mathit{KK}_k^{k+p}(|L|).
\end{equation*}
\end{prop}

\section{Improved upper bounds} \label{sterbounds}

The upper bound ${\it KK}$ on itself is neat and simple as it
takes as parameters only two numbers: the current size $k$, and
the number $|L|$ of current frequent patterns. However, in
reality, when we have arrived at a certain level $k$, we do not
merely have the cardinality: we have the actual set $L$ of current
$k$-patterns! For example, if the frequent patterns in the current
pass are all disjoint, our current upper bound will still estimate
their number to a certain non-zero figure. However, by the
pairwise disjointness, it is clear that no further patterns will
be possible at all. In sum, because we have richer information
than a mere cardinality, we should be able to get a better upper
bound.

To get inspiration, let us recall that
the candidate generation process of the Apriori algorithm works in
two steps.  In the \emph{join} step, we join $L$ with itself to
obtain a superset of $C_{k+1}$.  The
union $p \cup q$ of two patterns $p,q \in L$ is inserted in
$C_{k+1}$ if they share their $k-1$ smallest items:
\begin{tabbing}
{\bf insert into} $C_{k+1}$ \\
{\bf select} $p[1], p[2], \ldots, p[k], q[k]$ \\
{\bf from} $L_k$ $p$, $L_k$ $q$ \\
{\bf where} $p[1] = q[1]$, \dots, $p[k-1] = q[k-1]$, $p[k] < q[k]$
\end{tabbing}
Next, in the \emph{prune} step, we delete every pattern $c \in
C_{k+1}$ such that some $k$-subset of $c$ is not in $L$.

Let us now take a closer look at the join step from another point
of view. Consider a family of all frequent patterns of size $k$
that share their $k-1$ smallest items, and let its cardinality be $n$.
If we now remove from each of these patterns all these shared $k-1$ smallest
items, we get exactly $n$ distinct single-item patterns.  The number of pairs
that can be formed from these single items, being
$\binom{n}{2}$, is exactly the
number of candidates the join step will generate for the family under
consideration.  We thus get an obvious
upper bound on the total number of candidates by
taking the sum of all $\binom{n_f}{2}$, for every possible family $f$.

This obvious upper bound on $|C_{k+1}|$, which we denote by ${\it
obvious}_{k+1}(L)$, can be recursively computed in the following
manner. Let $I$ denote the set of items occurring in $L$. For an
arbitrary item $x$, define the set $L^x$ as $$ L^x = \{s - \{x\}
\mid s \in L \ {\rm and} \ x = \min s\}. $$  Then
\begin{equation*}
{\it obvious}_{k+1}(L) := \begin{cases}
\displaystyle \binom{|L|}{2} & \text{if $k=1$;} \\
\sum_{x \in I} {\it obvious}_k(L^x) & \text{if $k>1$.}
\end{cases}
\end{equation*}

This upper bound is much too crude, however, because it
does not take the prune step into account, only the join step.
The join step only
checks two $k$-subsets of a potential candidate
instead of all $k+1$
$k$-subsets.

However,
we can generalize this method such that more subsets
will be considered.  Indeed, instead of taking a family of all
frequent patterns sharing their $k-1$ smallest items, we can take all
frequent patterns sharing only their $k'$ smallest items, for some
$k' \leq k-1$.
If we then remove
these $k'$ shared items from each pattern in the family, we get
a new set $L'$ of $n$ patterns of size $k-k'$.  If we now consider the set
$C'$ of
candidates (of size $k-k'+1$) for $L'$, and
add back to each of them the previously removed $k'$ items, we obtain
a pruned set of candidates of size $k+1$, where instead of just two
(as in the join step), $k-k'+1$ of the $k$-subsets were checked in the
pruning.  Note that we can get the estimate $KK_{k-k'}^{k-k'+1}(|L'|)$
on the cardinality of $C'$ from our upper bound Theorem~\ref{kk}.

Doing this for all possible values of $k'$ yields an improved
upper bound on $|C_{k+1}|$, which we denote by ${\it
improved}_{k+1}(L)$, and which is computed by refining the
recursive procedure for the obvious upper bound as follows:
\begin{equation*} {\it improved}_{k+1}(L) := \begin{cases}
\displaystyle \binom{|L|}{2} & \text{if $k=1$;} \\
\min\{KK^{k+1}_k(|L|), \sum_{x \in I} {\it improved}_k(L^x)\} & \text{if $k>1$.}
\end{cases}
\end{equation*}

Actually, as in the previous section, we can do this not only to estimate
$|C_{k+1}|$, but also more generally to estimate $|C_{k+p}|$ for any $p \geq
1$.  Henceforth we will denote our general
improved upper bound by $KK^*_{k+p}(L)$. The general definition is as follows:
\begin{equation*}
{\it KK}^*_{k+p}(L) :=
\begin{cases}
\displaystyle {\it KK}_k^{k+p}(|L|) & \text{if $k=1$;} \\ \min \{
\mathit{KK}_{k}^{k+p}(|L|), \sum_{x \in I}
\mathit{KK}^*_{k+p-1}(L^x) \} & \text{if $k>1$.}
\end{cases}
\end{equation*} (For the base case, note that ${\it KK}_k^{k+p}(|L|)$, when $k=1$,
is nothing but $\binom{|L|}{p+1}$.)

By definition, ${\it KK}_{k+p}^*$ is always smaller than ${\it
KK}_{k}^{k+p}$. We now prove formally that it is still an upper
bound on the number of candidate patterns of size $k+p$:

\begin{thm} \label{stertheorema}
$$|C_{k+p}(L)| \leq {\it KK}^*_{k+p}(L).$$
\end{thm}
\begin{proof}
By induction on $k$. The base case $k=1$ is clear. For $k>1$, it
suffices to show that for all $p>0$
\begin{equation}
\label{toshow} C_{k+p}(L) \subseteq \bigcup_{x \in I}
C_{k+p-1}(L^x) + x.
\end{equation}
(For any set of patterns $H$, we denote $\{h \cup \{x\} \mid h \in H\}$ by
$H+x$.)

From the above containment we can conclude
\begin{align*}
|C_{k+p}(L)| & \leq |\bigcup_{x \in I} C_{k+p-1}(L^x) + x| \\
 & \leq \sum_{x \in I} |C_{k+p-1}(L^x) + x| \\
 & = \sum_{x \in I} |C_{k+p-1}(L^x)| \\
 & \leq \sum_{x \in I} {\it KK}^*_{k+p-1}(L^x)
\end{align*}
where the last inequality is by induction.

To show (\ref{toshow}), we need to show that for every $p>0$ and
every $s\in C_{k+p}(L)$, $s-\{x\}\in C_{k+p-1}(L^x)$, where
$x=\min s$. This means that every subset of
$s-\{x\}$ of size $k-1$ must be an element of
$L^x$. Let $s-\{x\}-\{y_1,\ldots,y_p\}$ be such a subset. This
subset is an element of $L^x$ iff $s-\{y_1,\ldots,y_p\}\in L$ and
$x=\min(s-\{y_1,\ldots,y_p\})$. The first condition follows from
$s\in C_{k+p}(L)$, and the second condition is trivial. Hence
the theorem.
\end{proof}

A natural question is why we must take the minimum in the
definition of ${\it KK}^*$.  The answer is that the two terms of
which we take the minimum are incomparable.  The example of an $L$
where all patterns are pairwise disjoint, already mentioned in the
beginning of this section, shows that, for example, ${\it
KK}_k^{k+1}(|L|)$ can be larger than the summation $\sum_{x \in I}
\mathit{KK}^*_k(L^x)$.  But the converse is also possible:
consider $L = \{\{1,2\},\{1,3\}\}$.  Then ${\it KK}_2^3(L)=0$, but
the summation yields 1.

\begin{exmp}
Let $L$ consist of $\{5,7,8\}$ and $\{5,8,9\}$ plus all $19$
$3$-subsets of $\{1,2,3,4,5\}$ and $\{3,4,5,6,7\}$. Because $21 =
\binom{6}{3} + \binom{2}{2},$ we have ${\it KK}_3^{4}(21)=15$,
${\it KK}_3^{5}(21)=6$ and ${\it KK}_3^{6}(21)=1.$ On the other
hand,

\begin{equation*}
\begin{split}
{\it KK}^*_4(L) & =
{\it KK}^*_3(L^1) + {\it KK}^*_3(L^2) + {\it KK}^*_3(L^3) + {\it
KK}^*_3(L^4) \\ & \quad + {\it KK}^*_2((L^5)^6) + {\it KK}^*_2((L^5)^7)
+ {\it KK}^*_2((L^5)^8) + {\it KK}^*_2((L^5)^9) \\ & \quad + {\it
KK}^*_3(L^6) + {\it KK}^*_3(L^7) + {\it KK}^*_3(L^8) + {\it
KK}^*_3(L^9)
\\ & = 4 + 1 + 4 + 1 + 0 + \cdots + 0 \\ & = 10
\end{split}
\end{equation*}
and
\begin{equation*}
\begin{split}
{\it KK}^*_5(L) & = {\it KK}^*_4(L^1) + {\it KK}^*_4(L^2) + {\it
KK}^*_4(L^3) + {\it KK}^*_4(L^4) \\ & \quad + {\it KK}^*_3((L^5)^6) +
{\it KK}^*_3((L^5)^7) + {\it KK}^*_3((L^5)^8) + {\it
KK}^*_3((L^5)^9) \\ & \quad + {\it KK}^*_4(L^6) + {\it KK}^*_4(L^7) +
{\it KK}^*_4(L^8) + {\it KK}^*_4(L^9) \\ & = 1 + 0 + 1 + 0 + 0 +
\cdots + 0 \\ & = 2.
\end{split}
\end{equation*}
Indeed, we have $10$ $4$-subsets of $\{1,2,3,4,5\}$ and
$\{3,4,5,6,7\},$ and the two $5$-sets themselves.
\end{exmp}

We can also improve the upper bound $\mu_k(|L|)$ on $\maxsize{L}$.
In analogy with Proposition~\ref{boils}, we define:
\begin{equation*}
\mu^*_k(L):=k+\min\{p\mid {\it KK}^*_{k+p}(L)=0\}-1.
\end{equation*}
We then have:
\begin{prop} \label{muster}
$$\maxsize{L} \leq \mu^*_k(L) \leq \mu_k(L).$$
\end{prop}

We finally use Theorem~\ref{stertheorema} for improving the upper
bound ${\it KK}_k^{\rm total}$ on the total number of candidate
patterns. We define:
\begin{equation*}
{\it KK}^*_{\rm total}(L):=\sum_{p \geq 1} 
{\it KK}^*_{k+p}(L).
\end{equation*}
Then we have:
\begin{prop}\label{totalster}
The total number of candidate patterns that can be generated  from a set
$L$
of $k$-patterns is
bounded by ${\it KK}^*_{\rm total}(L)$. Moreover,
$${\it KK}^*_{\rm total}(L)\leq {\it KK}_k^{\rm total}(L).$$
\end{prop}

\section{Generalized upper bounds} \label{generalbounds}

The upper bounds presented in the previous sections work well for
algorithms that generate and test candidate patterns of one
specific size at a time.  However, a lot of algorithms generate
and test patterns of different sizes within the same pass of the
algorithm~\cite{dic,maxminer,sampling}. Hence, these algorithms
know in advance that several patterns of size larger than $k$
are frequent or not. Since our upper bound is solely based on the
patterns of a certain length $k$, it does not use information
about patterns of length larger than $k$.

Nevertheless, these larger sets could give crucial information.
More specifically, suppose we have generated all frequent patterns
of size $k$, and we also already know in advance that a certain
set of size larger than $k$ is not frequent.  Our upper bound on
the total number of candidate patterns that can still be
generated, would disregard this information. We will therefore
generalize our upper bound such that it will also incorporate this
additional information.

\subsection{Generalized ${\it KK}$-bound}

From now on, $L$ is some family of sets of patterns
$L_k, L_{k+1}, \ldots, L_{k+q}$ which are known to be frequent,
such that $L_{k+p}$ contains patterns of size $k+p$, and all
$k+p-1$-subsets of all patterns in $L_{k+p}$ are in $L_{k+p-1}$.
We denote by $|L|$ the sequence of numbers $|L_k|,
|L_{k+1}|, \ldots, |L_{k+q}|$.

Similarly, let $I$ be a family of sets of patterns
$I_{k}, I_{k+1}, \ldots, I_{k+q}$ which are known to be
infrequent, such that $I_{k+p}$ contains patterns of size $k+p$
and all $k+p-1$-subsets of all patterns in $I_{k+p}$ are in
$L_{k+p-1}$. We denote by $|I|$ the sequence of numbers
$|I_{k}|,|I_{k+1}|,\ldots, |I_{k+q}|$. Note that for each $p \geq
0$, $L_{k+p}$ and $I_{k+p}$ are disjoint.

Before we present the general upper bounds, we also generalize our
notion of a candidate pattern.
\begin{defin}
A \emph{candidate pattern for $(L,I)$} of size
$k+p$ is a pattern which is not in $L_{k+p}$ or $I_{k+p}$, all of
its $k$-subsets are in $L_k$, and none of its subsets of size
larger than $k$ is included in $I_{k} \cup I_{k+1} \cup \cdots
\cup I_{k+q}$. For a given $p$, we denote the set of all
$k+p$-size candidate patterns for $(L,I)$ by
$C_{k+p}(L,I)$.
\end{defin}

We note:
\begin{lem} \label{lem10}
\begin{equation*}
C_{k+p}(L,I) =
\begin{cases}
C_{k+1}(L_k)\setminus (L_{k+1} \cup I_{k+1}) & \text{if $p=1$;} \\
C_{k+p}\bigl(C_{k+p-1}(L,I)\cup
L_{k+p-1}\bigr)\setminus (L_{k+p} \cup I_{k+p})
 & \text{if $p>1$.}
\end{cases}
\end{equation*}
\end{lem}
\begin{proof}
The case $p=1$ is clear. For $p>1$,
we show the inclusion in both directions.
\begin{itemize}

\item[$\supseteq$]
For every set in $C_{k+p}\bigl(C_{k+p-1}(L,I)\cup
L_{k+p-1}\bigr)$, we know that all of its $k$-subsets are always contained
in a $k+p-1$ subset, and these are in $C_{k+p-1}(L,I) \cup L_{k+p-1}$.
By definition, we know that for every set in $C_{k+p-1}(L,I)$,
all of its $k$-subsets are in $L_k$.
Also, for every set in $L_{k+p-1}$, all of its $k$-subsets are in $L_k$.
By definition, for every set in $C_{k+p-1}(L,I)$,
all of its $k+p-i$-subsets are not in $I_{k+p-i}$.
Also, for every set in $L_{k+p-1}$,
all of its $k+p-i$-subsets are in $L_{k+p-i}$ and hence they are not in $I_{k+p-i}$
since they are disjoint.
By definition, none of the patterns in $L_{k+p} \cup I_{k+p}$ are in
$C_{k+p}(L,I)$.

\item[$\subseteq$] It suffices to show that for every set in
$C_{k+p}(L,I)$, every $k+p-1$-subset $s$ is in
$C_{k+p-1}(L,I)\cup L_{k+p-1}$. Obviously, this
is true, since if it is not already in $L_{k+p-1}$, still all
$k$-subsets of $s$ must be in $L_k$, $s$ can not be in $I_{k+p-1}$
and none of its subsets can be in any $I_{k+p-\ell}$ with
$\ell>1$.
\end{itemize}
\end{proof}

Hence, we define
\begin{multline*}
\mathit{gKK}^{k+p}_k(|L|,|I|) :=\\
\begin{cases}
\mathit{KK}^{k+1}_k(|L_k|) - |L_{k+1}|-|I_{k+1}| & \text{if $p=1$;} \\
\mathit{KK}^{k+p}_{k+p-1}(\mathit{gKK}^{k+p-1}_k(|L|,|I|) + |L_{k+p-1}|) - |L_{k+p}|-|I_{k+p}|
& \text{if $p>1$,}
\end{cases}
\end{multline*}
and obtain:
\begin{thm} \label{gkkthm}
$$|C_{k+p}(L,I)| \leq  \mathit{gKK}^{k+p}_k(|L|,|I|) \leq \mathit{KK}_k^{k+p}(|L_k|) - |L_{k+p}| - |I_{k+p}|.$$
\end{thm}
\begin{proof}
The first inequality is clear by Lemma~\ref{lem10}. The second inequality is by induction on $p$.
The base case $p=1$ is by definition. For $p>1$, we have:
\begin{align*}
\mathit{gKK}^{k+p}_k(|L|,|I|)
& = \mathit{KK}^{k+p}_{k+p-1}(\mathit{gKK}^{k+p-1}_k(|L|,|I|)+ |L_{k+p-1}|) \\
& \phantom{=} - |L_{k+p}| - |I_{k+p}| \\
& \leq \mathit{KK}^{k+p}_{k+p-1}(\mathit{KK}_k^{k+p-1}(|L_k|) - |I_{k+p-1}|) - |L_{k+p}| - |I_{k+p}|\\
& \leq \mathit{KK}^{k+p}_{k+p-1}(\mathit{KK}_k^{k+p-1}(|L_k|)) - |L_{k+p}| - |I_{k+p}|\\
& = \mathit{KK}_k^{k+p}(|L_k|) - |L_{k+p}| - |I_{k+p}|
\end{align*}
where the first inequality is by induction and because of the
monotonicity of $\mathit{KK}$, the second inequality also because
of the monotonicity of $\mathit{KK}$ and the last equality follows
from
$$ \mathit{KK}_k^{k+p}(|L_k|))  = \mathit{KK}^{k+p}_{k+p-1}(\mathit{KK}_k^{k+p-1}(|L_k|)).$$
\end{proof}

Again, we can also generalize the upper bound on the maximal size
of a candidate pattern, denoted by $\maxsize{L,I}$, and the upper
bound on the total number of candidate patterns, both also
incorporating $(L,I)$:
\begin{equation*}
g\mu(|L|,|I|):=k+\min\{p\mid
\mathit{gKK}_k^{k+p}(|L|,|I|)=0\}-1
\end{equation*}
\begin{equation*}
\mathit{gKK}_k^{\rm total}(|L|,|I|):=\sum_{p\geq
1} \mathit{gKK}_k^{k+p}(|L|,|I|).
\end{equation*}
We obtain:
\begin{prop}
$$\maxsize{L,I} \leq g\mu(|L|,|I|) \leq \mu(|L|).$$
\end{prop}
\begin{prop}
The total number of candidate patterns that can be generated from
$(L,I)$ is bounded by $\mathit{gKK}_k^{\rm
total}(|L|,|I|)$. Moreover, $$\mathit{gKK}_k^{\rm
total}(|L|,|I|)\leq \mathit{KK}_k^{\rm
total}(|L_k|).$$
\end{prop}

\begin{exmp}
Suppose $L_3$ consists of all subsets of size $3$ of
the set $\{1,2,3,4,\allowbreak 5,6\}$. Now assume we already know that
$I_4$ contains patterns $\{1,2,3,4\}$ and $\{3,4,5,6\}$. The ${\it
KK}$ upper bound presented in the previous section would estimate
the number of candidate patterns of sizes $4, 5$, and $6$ to be at
most $\binom{6}{4}=15$, $\binom{6}{5}=6$, and $\binom{6}{6}=1$
respectively. Nevertheless, using the additional information,
$\mathit{gKK}$ can already reduce these numbers to
$13, 3$, and $0$. Also, $\mu$ would predict the maximal size of a
candidate pattern to be $6$, while $g\mu$ can already predict this number
to be at most $5$. Similarly, $\mathit{KK}_{\rm total}$ would predict
the total number of candidate patterns that can still be generated
to be at most $22$, while $\mathit{gKK}_{\rm total}$ can already
deduce this number to be at most $16$.
\end{exmp}

\subsection{Generalized ${\it KK}^*$-bound}

Using the generalized basic upper bound, we can now also
generalize our improved upper bound $\mathit{KK}^*$. For an
arbitrary item $x$, define the family of sets $L^x$ as
$L_k^x, L_{k+1}^x, \ldots, L_{k+q}^x$, and $I^x$ as
$I_k^x, I_{k+1}^x, \ldots, I_{k+q}^x$. We define:
\begin{multline*} \label{gkkster}
\mathit{gKK}^*_{k+p}(L,I) := \\
\begin{cases}
\displaystyle \mathit{gKK}_k^{k+p}(|L|,|I|) & \text{if $k=1$;} \\
\min \{ \mathit{gKK}_{k}^{k+p}(|L|,|I|), \sum_{x
\in I} \mathit{gKK}^*_{k+p-1}(L^x,I^x) \} &
\text{if $k>1$.}
\end{cases}
\end{multline*}
We then have:
\begin{thm} \label{gkksterthm}
$$|C_{k+p}(L,I)| \leq \mathit{gKK}^*_{k+p}(L,I) \leq
\mathit{KK}^*_{k+p}(L_k)- |L_{k+p}|-|I_{k+p}|.$$
\end{thm}
\begin{proof}
The proof of the first inequality is similar to the proof of Theorem~\ref{stertheorema},
instead that we now need to show that for all $p>0$,
$$C_{k+p}(L,I) \subseteq \bigcup_{x\in I} C_{k+p-1}(L^x,I^x)+x.$$
Therefore, we need to show for every $s \in C_{k+p}(L,I)$,
$s-\{x\} \in C_{k+p-1}(L^x,I^x)$, where $x = \min s$. First, this
means that every subset of $s-\{x\}$ of size $k-1$ must be in
$L^x_k$. Let $s-\{x\}-\{y_1,\ldots,y_p\}$ be such a subset. This
subset is an element of $L^x_k$ if and only if
$s-\{y_1,\ldots,y_p\}\in L_k$ and $x=\min(s-\{y_1,\ldots,y_p\})$.
The first condition follows from $s\in C_{k+p}(L,I)$, and the
second condition is trivial. Second, we need to show that
$s-\{x\}$ is not in $L_{k+p}^x$. Since $s\in C_{k+p}(L,I)$,  $s$
is not in $L_{k+p}$ and hence $s-\{x\}$ cannot be in $L_{k+p}^x$.
Finally,  we need to show that none of the subsets of $s-\{x\}$ of
size greater than $k-1$ are in $I_{k+1}^x,\ldots,I_{k+p-1}^x$. Let
$s-\{x\}-\{y_1,\ldots,y_m\}$ be such a subset. Since  $s\in
C_{k+p}(L,I)$, $s-\{y_1,\ldots,y_m\}$ is not in $I_{k+p-m}$, and
hence $s-\{x\}-\{y_1,\ldots,y_m\}$ cannot be in $I_{k+p-m}^x$.

We prove the second inequality by induction on $k$. The base case $k=1$ is clear.
For all $k>0$, we have
\begin{align*}
& \mathit{gKK}^*_{k+p}(L,I) \\
& = \min \{ \mathit{gKK}_{k}^{k+p}(|L|,|I|), \sum_{x
\in I} \mathit{gKK}^*_{k+p-1}(L^x,I^x) \} \\
& \leq \min \{ \mathit{KK}_{k}^{k+p}(|L_k|) - |L_{k+p}| - |I_{k+p}| ,
\sum_{x \in I} \mathit{KK}^*_{k+p-1}(L^x_k) - |L_{k+p}^x| - |I_{k+p}^x| \} \\
& = \min \{\mathit{KK}_{k}^{k+p}(|L|), \sum_{x \in I}\mathit{KK}^*_{k+p-1}(L^x) \} - |L_{k+p}| - |I_{k+p}| \\
& = \mathit{KK}^*_{k+p}(L_k)- |L_{k+p}|-|I_{k+p}|
\end{align*}
where the left hand side of the minimum in the inequality is by Theorem~\ref{gkkthm} and the
right hand side is by induction.
\end{proof}

Again, we get an upper bound on maxsize($L,I$):
\begin{equation*}
g\mu^*(L,I):=k+\min\{p\mid
\mathit{gKK}^*_{k+p}(L,I)=0\}-1,
\end{equation*}
and on the total number of candidate patterns that can still be
generated:
\begin{equation*}
\mathit{gKK}^*_{\rm total}(L,I):=\sum_{p\geq 1} \mathit{gKK}^*_{k+p}(L,I).
\end{equation*}
We then have the following analogous propositions to~\ref{muster} and~\ref{totalster}:
\begin{prop}
$$\maxsize{L,I} \leq g\mu^*(L,I) \leq \mu^*(L).$$
\end{prop}
\begin{prop}
The total number of candidate patterns that can be generated  from
$(L,I)$ is bounded by $\mathit{gKK}^*_{\rm
total}(L,I)$. Moreover, $$\mathit{gKK}^*_{\rm
total}(L,I)\leq \mathit{KK}^*_{\rm
total}(L_k).$$
\end{prop}

\begin{exmp}
Consider the same set of patterns as in the previous example. I.e.,
$L_{3}$ consists of all subsets of size $3$ of the set
$\{1,2,3,4,5,6\}$ and $\{1,2,3,4\}$ and $\{3,4,5,6\}$ are
included in $I_4$. The ${\it KK}^*$
upper bound presented in the previous section would also estimate the
number of candidate patterns of sizes $4, 5$, and $6$ to be at
most $\binom{6}{4}=15$, $\binom{6}{5}=6$, and $\binom{6}{6}=1$
respectively. Nevertheless, using the additional information,
$\mathit{gKK}^*$ can perfectly predict these numbers to be $13, 2$,
 and $0$. Again, $\mu^*$ would predict the maximal size of a
candidate pattern to be $6$, while $g\mu^*$ can already predict this number
to be at most $5$. Similarly, $\mathit{KK}^*_{\rm total}$ would predict
the total number of candidate patterns that can still be generated
to be at most $22$, while $\mathit{gKK}^*_{\rm total}$ can already
deduce this number to be at most $15$.
\end{exmp}

\section{Efficient Implementation} \label{implement}

For simplicity reasons, we will restrict ourselves to the
explanation of how the improved upper bounds can be implemented.
The proposed implementation can be easily extended to support the
computation of the general upper bounds.

To evaluate our upper bounds we implemented an optimized version
of the Apriori algorithm using a trie data structure to store all
generated patterns, similar to the one described by Brin et
al.~\cite{dic}. This trie structure makes it cheap and
straightforward to implement the computation of all upper bounds.
Indeed, a top-level subtrie (rooted at some singleton pattern
$\{x\}$) represents exactly the set $L^x$ we defined in
Section~\ref{sterbounds}. Every top-level subtrie of this subtrie
(rooted at some two-element pattern $\{x,y\}$) then represents
$(L^x)^y$, and so on. Hence, we can compute the recursive bounds
while traversing the trie, after the frequencies of all candidate
patterns are counted, and we have to traverse the trie once more
to remove all candidate patterns that turned out to be infrequent.
This can be done as follows.

Remember, at that point, we have the current set of frequent
patterns of size $k$ stored in the trie.  For every node at depth
$d$ smaller than $k$, we compute the $k-d$-canonical
representation of the number of descendants this node has at depth
$k$, which can be used to compute $\mu_{k-d}$
(cf.~Proposition~\ref{muprop}), ${\it KK}_{k-d}^{\ell}$ for any
$\ell \leq \mu_{k-d}$ (cf.~Theorem~\ref{kk}) and hence also ${\it
KK}_{k-d}^{\rm total}$ (cf.~Proposition~\ref{kktotal}). For every
node at depth $k-1$, its ${\it KK}^*$ and $\mu^*$ values are equal
to its ${\it KK}$ and $\mu$ values respectively.  Then compute for
every $p > 0$, the sum of the ${\it KK}^*_{k-d+p-1}$ values of all
its children, and let ${\it KK}^*_{k-d+p}$ be the smallest of this
sum and ${\it KK}^{k-d+p}_{k-d}$ until this minimum becomes zero,
which also gives us the value of $\mu^*$. Finally, we can compute
${\it KK}^*_{\rm total}$ for this node. If this is done for every
node, traversed in a depth-first manner, then finally the root
node will contain the upper bounds on the number of candidate
patterns that can still be generated, and on the maximum size of
any such pattern. The soundness and completeness of this method
follows directly from the theorems and propositions of the
previous sections.

We should also point out that, since the numbers involved can
become exponentially large (in the number of items), an
implementation should take care to use arbitrary-length integers
such as provided by standard mathematical packages. Since the
length of an integer is only logarithmic in its value, the lengths
of the numbers involved will remain polynomially bounded.

\section{Experimental Evaluation} \label{experiment}

All experiments were performed on a 400MHz Sun Ultra Sparc with 512~MB
main memory, running Sun Solaris~8.  The algorithm was implemented in {C++}
and uses the {GNU MP} library for arbitrary-length integers~\cite{gmp}.

\paragraph{Data sets}
We have experimented using three real data sets, of which two are
publicly available, and one synthetic data set generated by the
program provided by the Quest research group at IBM
Almaden~\cite{datagenerator}. The mushroom data set contains
characteristics of various species of mushrooms, and was
originally obtained from the {UCI} repository of machine learning
databases~\cite{ucimlr}. The BMS-WebView-1 data set contains
several months worth of clickstream data from an e-commerce web
site, and is made publicly available by Blue Martini
Software~\cite{kddcup2000}. The basket data set contains
transactions from a Belgian retail store, but can unfortunately
not be made publicly available. Table~\ref{database} shows
the number of items and the number of transactions in each data
set. The table additionally shows the minimal support threshold we
used in our experiments for each data set, together with the
resulting number of iterations and the time (in seconds) which the
Apriori algorithm needed to find all frequent patterns.
\begin{table}
\centering
\begin{tabular}{|l|c|c|c|c|c|}
  \hline
  Data set & \#Items & \#Transactions & MinSup & \#It's & Time \\
  \hline
    T40I10D100K & 1\,000 & 100\,000 & 700 & 18 & 1\,700s \\
    mushroom & 120 & 8\,124 & 813 & 16 & 663s \\
    BMS-Webview-1 & 498 & 59\,602 & 36 & 15 & 86s \\
    basket & 13\,103 & 41\,373 & 5 & 11 & 43s \\
  \hline
\end{tabular}
\caption{Database Characteristics}
\label{database}
\end{table}

The results from the experiment with the real data sets were
not immediately as good as the results from the synthetic data
set. The reason for this, however, turned out to be the bad
ordering of the items, as explained next.

\paragraph{Reordering}
From the form of $L^x$, it can be seen that the order of the items
can affect the recursive upper bounds.
By computing the upper bound only for a subset of all frequent
patterns (namely $L^x$), we win by incorporating the structure of
the current collection of frequent patterns, but we also lose some
information.
Indeed, whenever we recursively restrict ourselves to a subtrie
$L^x$, then for every candidate pattern $s$ with $x = \min s$, we
lose the information about exactly one subpattern in $L$, namely
$s-{x}$.

We therefore would like to make it likely that many of these
excluded patterns are frequent. A good heuristic, which has
already been used for several other optimizations in frequent
pattern mining \cite{maxminer,dic, treeproject}, is to force the
most frequent items to appear in the most candidate patterns, by
reordering the single item patterns in increasing order of
frequency.

After reordering the items in the real life data set, using this
heuristic, the results became very analogous with the results
using the synthetic datasets.

\begin{figure}
\begin{center}
\includegraphics{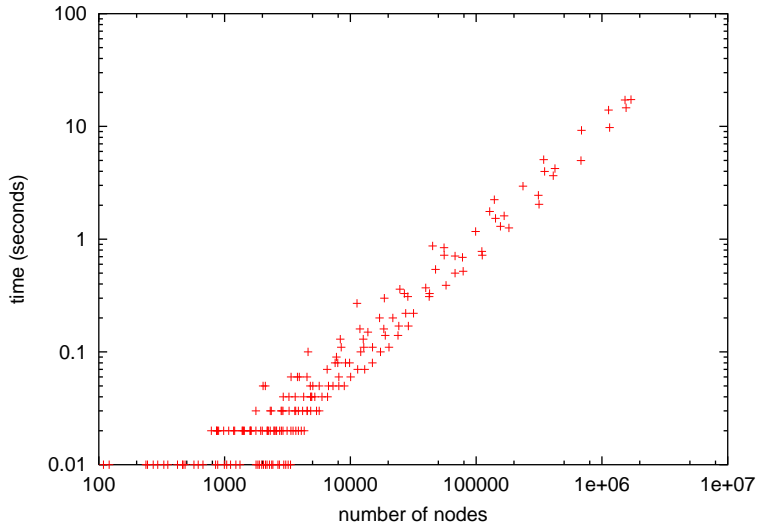}
\caption{Time needed to compute upper bounds is linear in the number of nodes.}
\label{timefig}
\end{center}
\end{figure}

\paragraph{Efficiency}
The cost for the computation of the upper bounds is negligible
compared to the cost of the complete algorithm. Indeed, the time
$T$ needed to calculate the upper bounds is largely dictated by
the number $n$ of currently known frequent sets. We have shown
experimentally that $T$ scales linearly with $n$. Moreover, the
constant factor in our implementation is very small (around
$0.00001$). We ran several experiments using the different data
sets and varying minimal support thresholds.  After every pass of
the algorithm, we registered the number of known frequent sets and
the time spent to compute all upper bounds, resulting in $145$
different data points. Figure~\ref{timefig} shows these results.

\paragraph{Upper bounds}

\begin{itemize}

\item Figure~\ref{fig:kknext} shows, after each level $k$, the
computed upper bound ${\it KK}$ and improved upper bound ${\it
KK}^*$ for the number of candidate patterns of size $k+1$, as well
as the actual number $|C_{k+1}|$ it turned out to be. We omitted
the upper bound for $k+1=2$, since the upper bound on the number
of candidate patterns of size $2$ is simply $\binom{|L|}{2}$, with
$|L|$ the number of frequent items.
\begin{figure}
\centering
\subfigure[basket]{\includegraphics{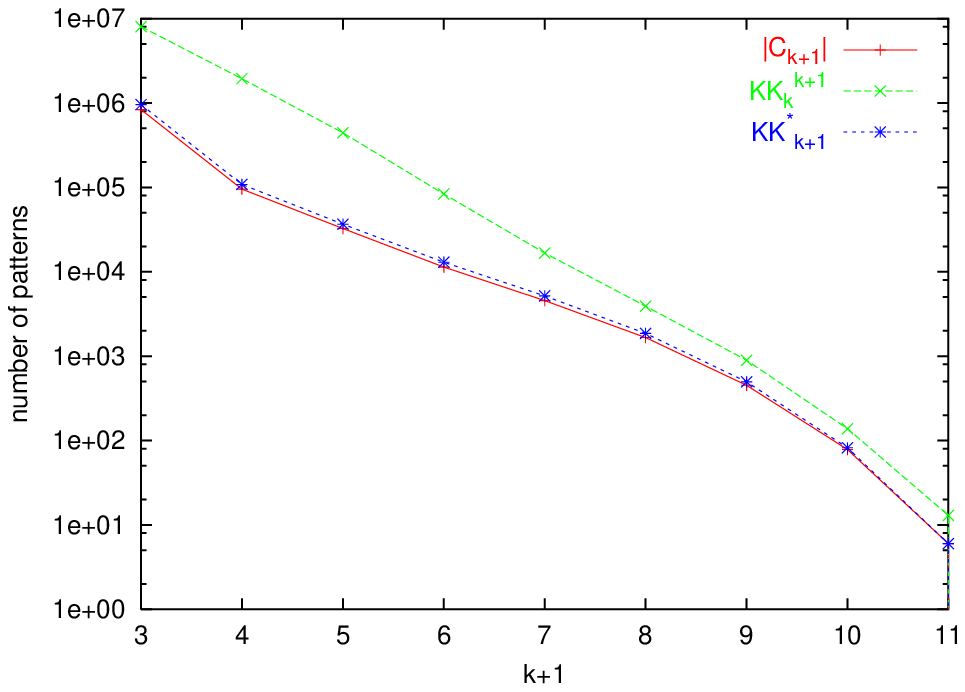}}\\
\subfigure[BMS-Webview-1]{\includegraphics{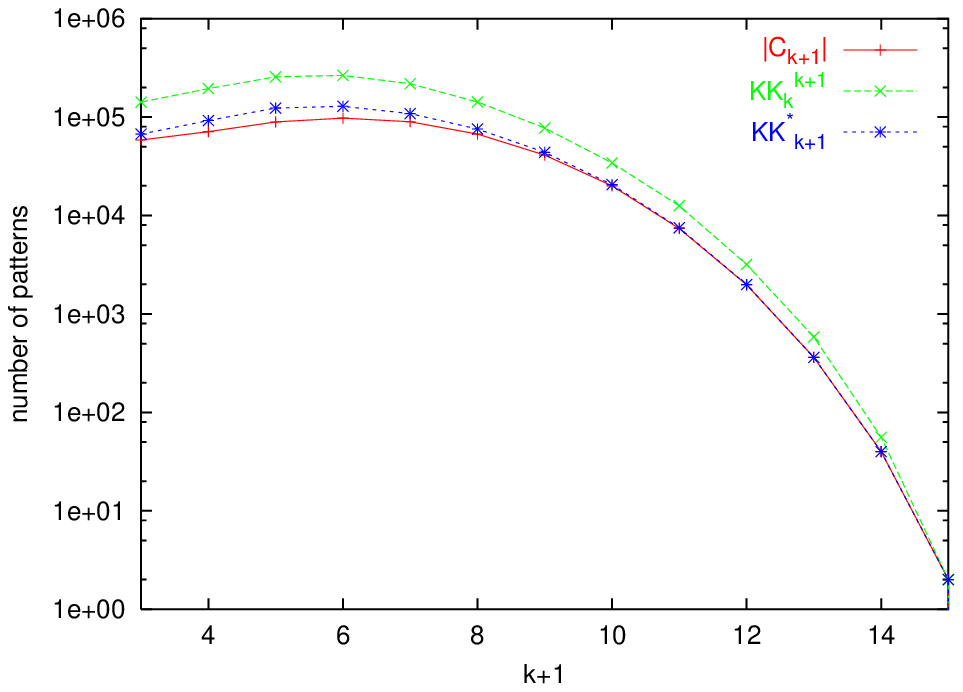}}
\caption{Actual and estimated number of candidate patterns.}
\label{fig:kknext}
\end{figure}
\addtocounter{figure}{-1}
\begin{figure}
\addtocounter{subfigure}{2}
\centering
\subfigure[T40I10D100K]{\includegraphics{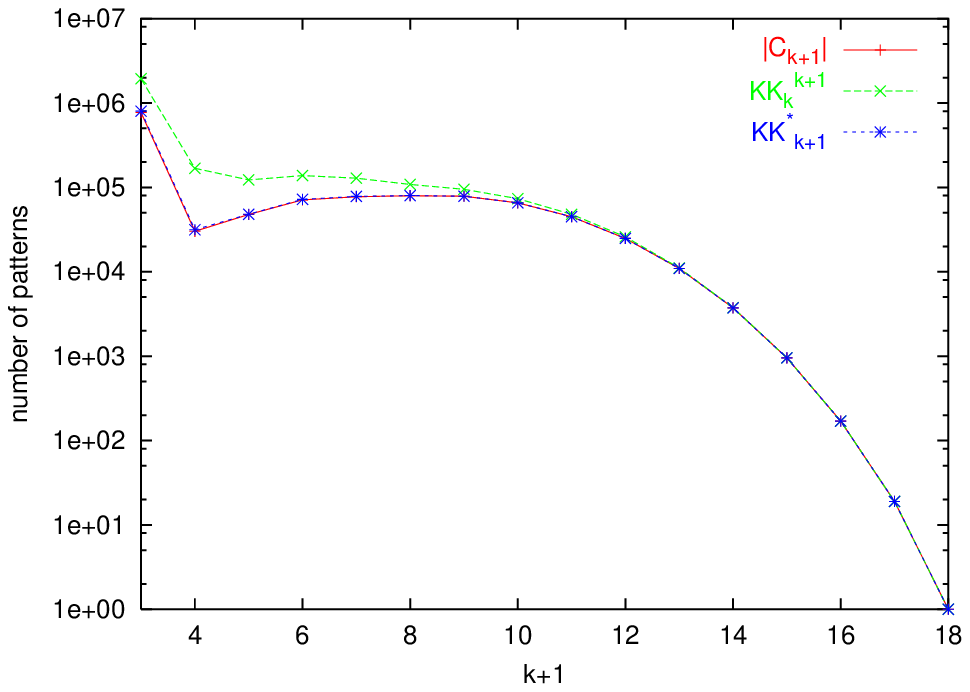}}\\
\subfigure[mushroom]{\includegraphics{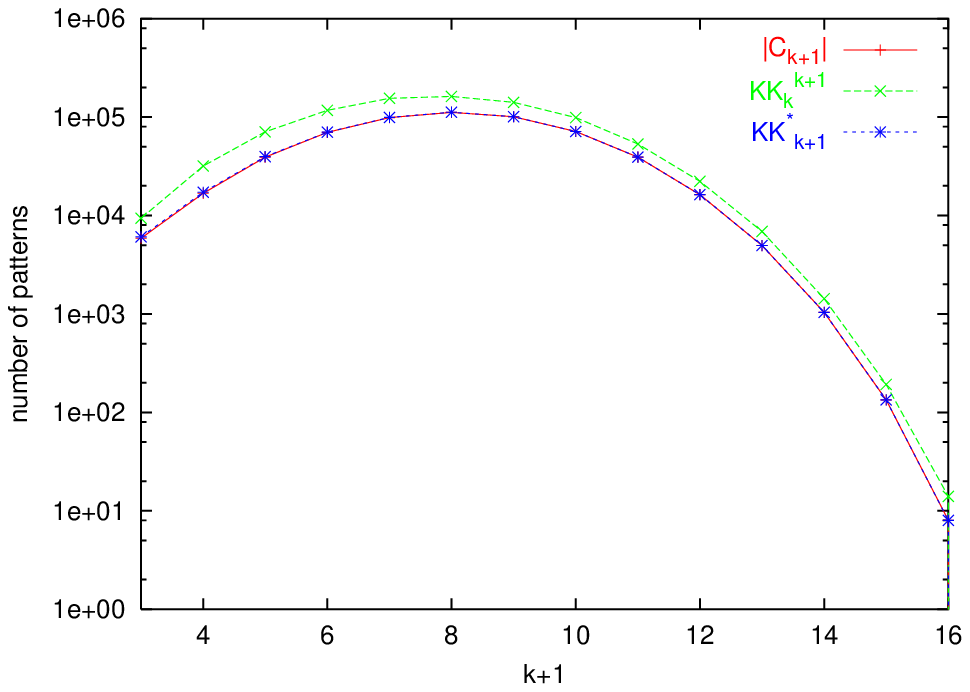}}
\caption{Actual and estimated number of candidate patterns.}
\end{figure}

\item Figure~\ref{fig:kktot} shows the upper bounds on the total number
of candidate patterns that could still be generated, compared to
the actual number of candidate patterns, $|C_{\mathrm{total}}|$, that
were effectively generated.
Again, we omitted the upper bound for $k=1$, since this number is simply $2^{|L|}-|L|-1$,
with $|L|$ the number of frequent items.
\begin{figure}
\centering
\subfigure[basket]{\includegraphics{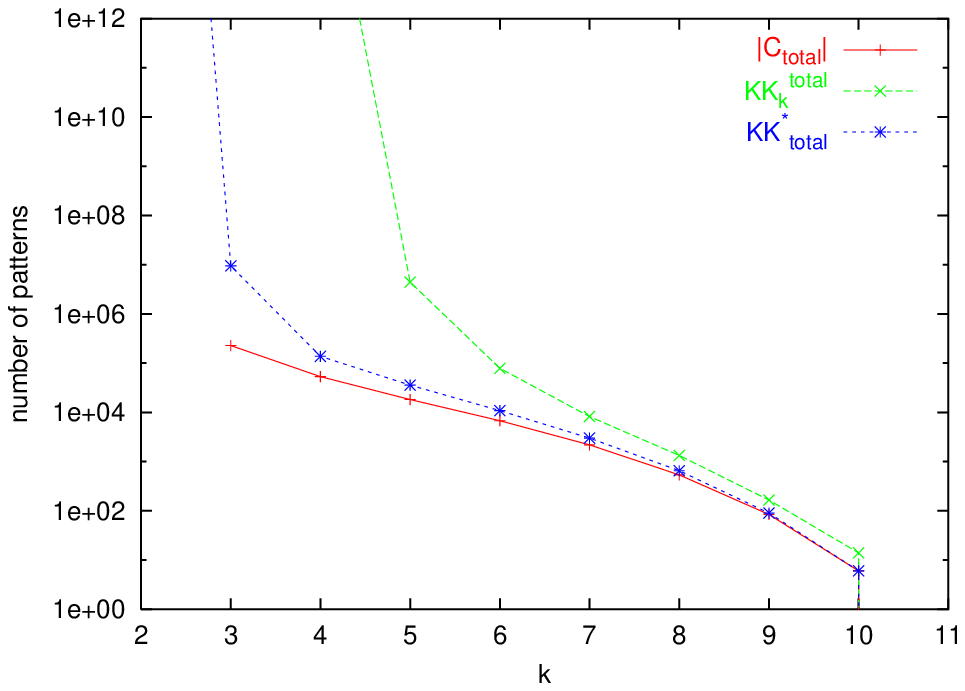}}\\
\subfigure[BMS-Webview-1]{\includegraphics{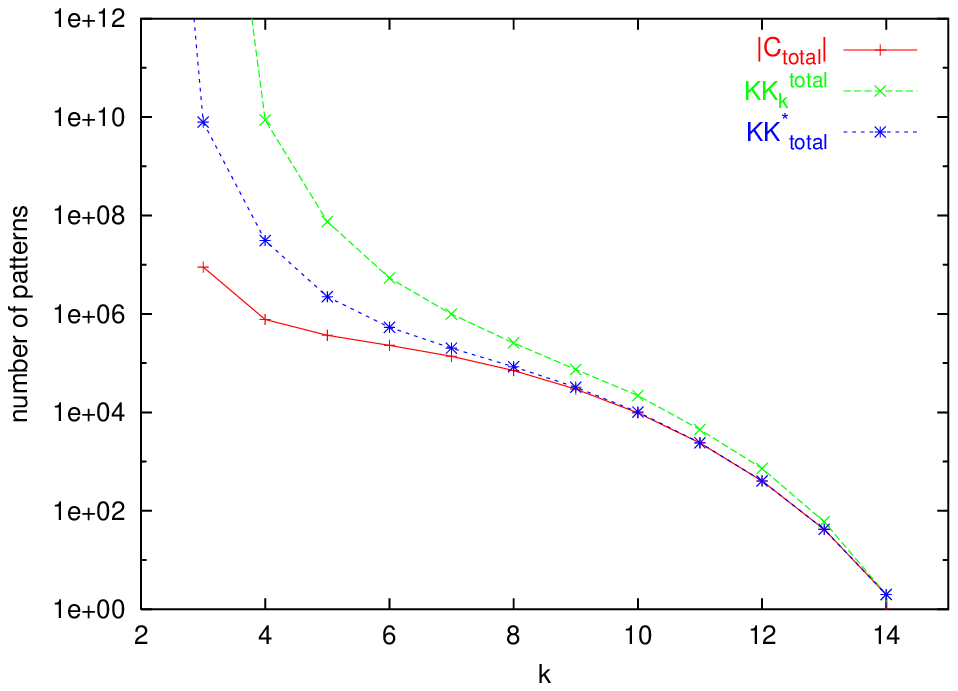}}
\caption{Actual and estimated total number of future candidate patterns.}
\label{fig:kktot}
\end{figure}
\addtocounter{figure}{-1}
\begin{figure}
\addtocounter{subfigure}{2}
\centering
\subfigure[T40I10D100K]{\includegraphics{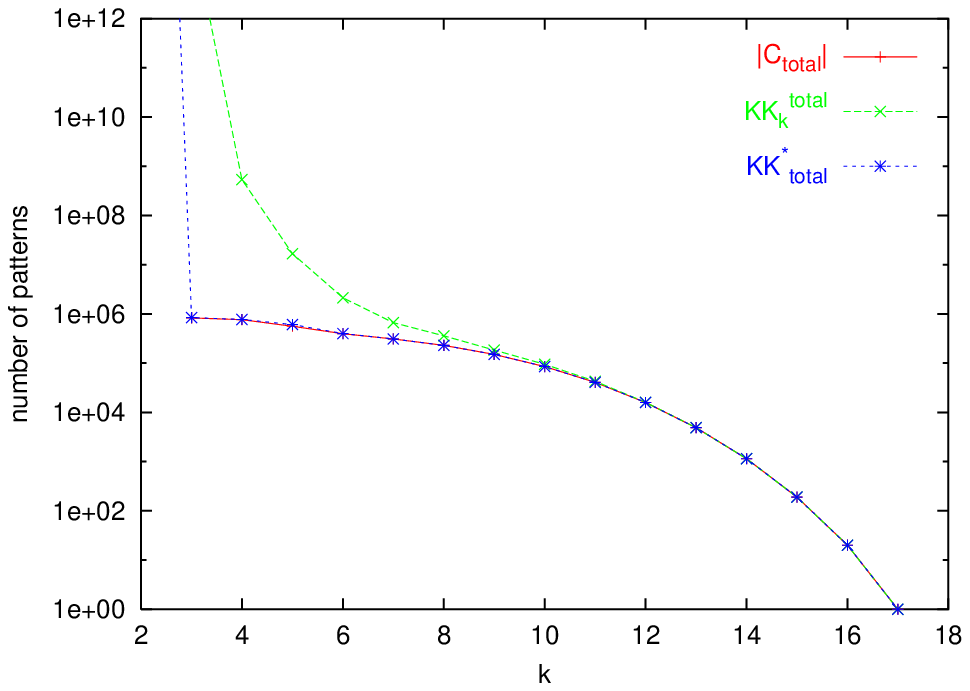}}\\
\subfigure[mushroom]{\includegraphics{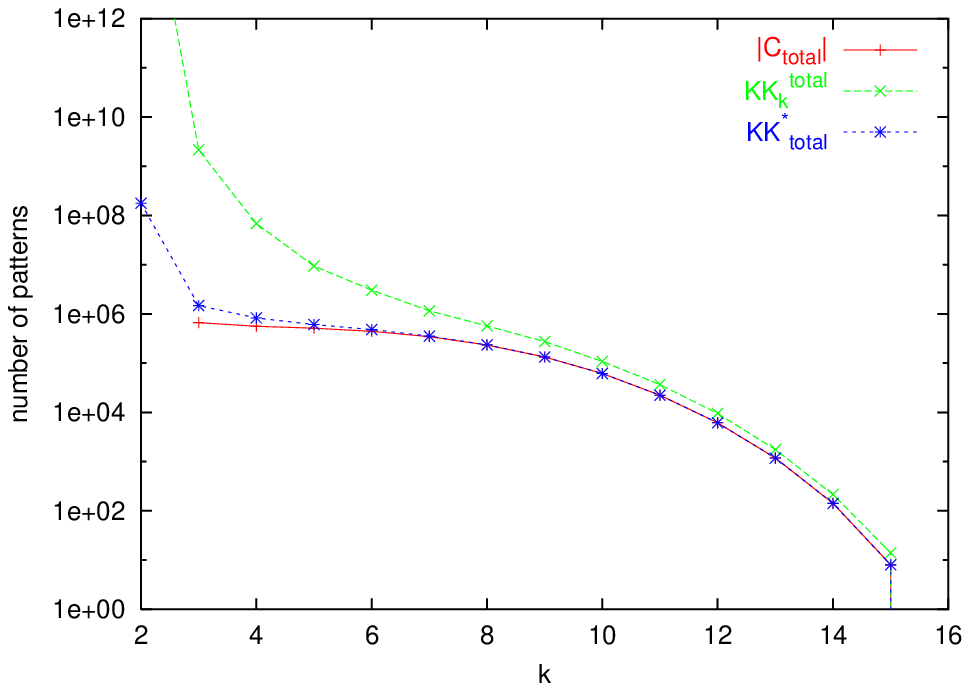}}
\caption{Actual and estimated total number of future candidate patterns.}
\end{figure}

\item Figure~\ref{fig:kkms} shows the computed upper bounds $\mu$ and
$\mu^*$ on the maximal size of a candidate pattern. Also here we omitted the result
for $k=1$, since this number is exactly the number of frequent items.
\begin{figure}
\centering
\subfigure[basket]{\includegraphics{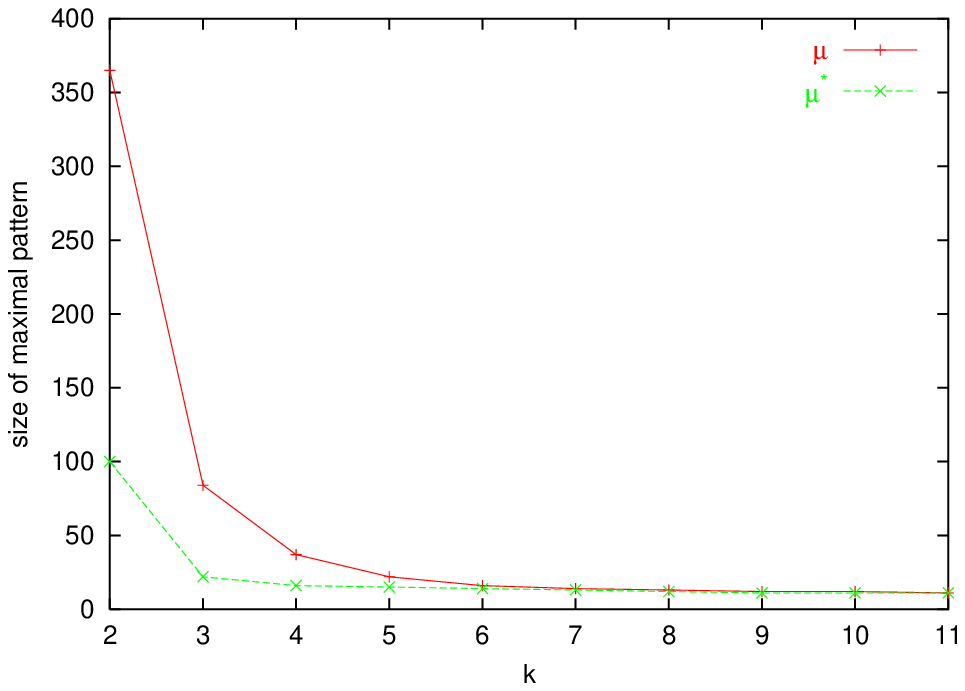}}\\
\subfigure[BMS-Webview-1]{\includegraphics{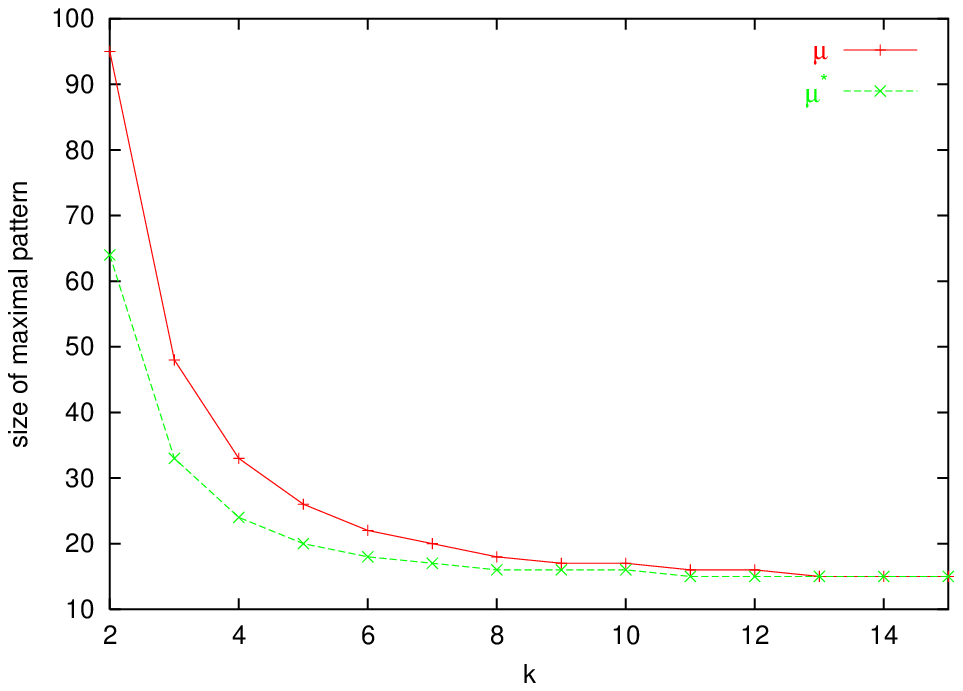}}
\caption{Estimated size of the largest possible candidate pattern.}
\label{fig:kkms}
\end{figure}
\addtocounter{figure}{-1}
\begin{figure}
\addtocounter{subfigure}{2}
\centering
\subfigure[T40I10D100K]{\includegraphics{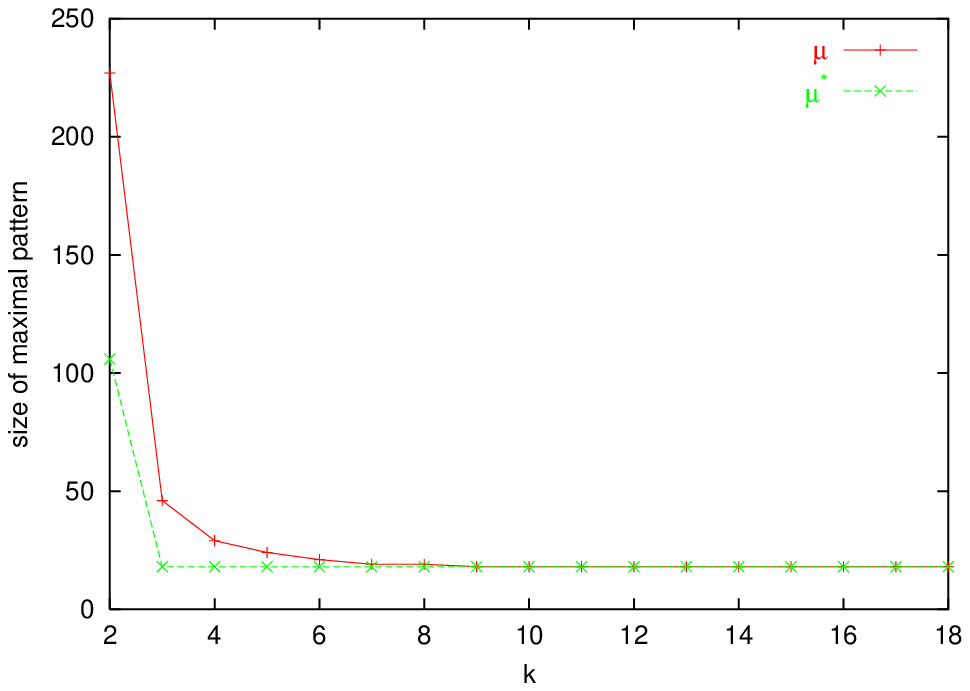}}\\
\subfigure[mushroom]{\includegraphics{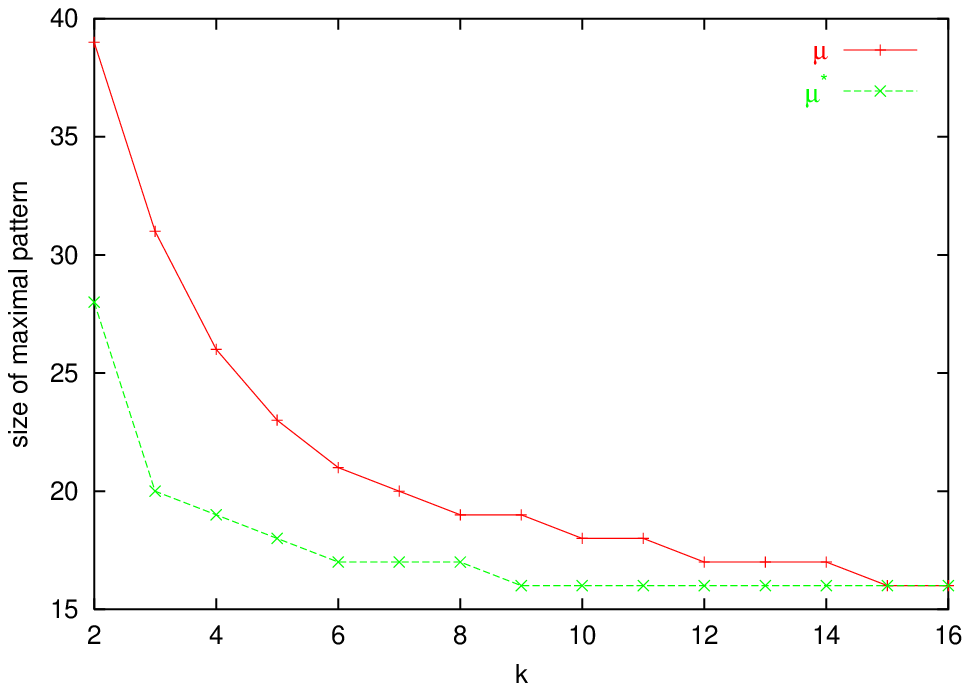}}
\caption{Estimated size of the largest possible candidate pattern.}
\end{figure}

\end{itemize}

The results are pleasantly surprising:
\begin{itemize}

\item Note that the
improvement of ${\it KK}^*$ over ${\it KK}$, and of $\mu^*$ over
$\mu$, anticipated by our theoretical discussion, is indeed
dramatic.

\item Comparing the computed upper bounds with the
actual numbers, we observe the high accuracy of the estimations
given by ${\it KK}^*$.  Indeed, the estimations of ${\it
KK}^*_{k+1}$ match almost exactly the actual number of candidate
patterns that has been generated at level $k+1$.
Also note that the number of candidate patterns in T40I10D100K is
decreasing in the first four iterations and then increases again.
This perfectly illustrates that the heuristic used for AprioriHybrid,
as explained in the related work section, would not work on this data set.
Indeed, any algorithm that exploits the fact that the current number of candidate patterns is
small enough and there were fewer candidate patterns in the
current iteration than in the previous iteration, would falsely interpret
these observations, since the number of candidate patterns in the next iterations
increases again.  The presented upper bounds perfectly predict this increase.

\item The upper bounds
on the total number of candidate patterns are still very large
when estimated in the first few passes, which is not surprising
because at these initial stages, there is not much information
yet. For the mushroom and the artificial data sets, the upper bound
is almost exact when the frequent patterns of size $3$ are known.
For the basket data set, this result is obtained when the frequent patterns
of size $4$ are known and size $6$ for the BMS-Webview-1 data set.

\item We also performed experiments for varying minimal support
thresholds. The results obtained from these experiments were
entirely similar to those presented above.

\end{itemize}

\paragraph{Combining iterations}

As discussed in the Introduction, the proposed upper bound can be
used to protect several improvements of the Apriori algorithm from
generating too many candidate patterns.  One such improvement
tries to combine as many iterations as possible in the end, when
only few candidate patterns can still be generated. We have
implemented this technique within our implementation of the
Apriori algorithm.

\begin{figure}
\centering
\subfigure[basket]{\includegraphics{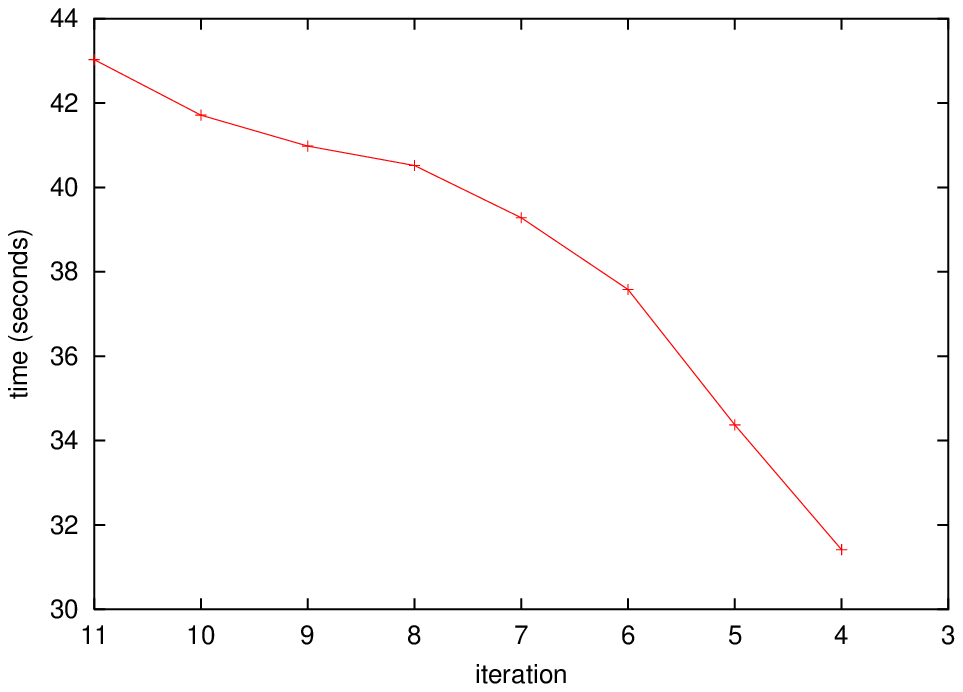}}\\
\subfigure[BMS-Webview-1]{\includegraphics{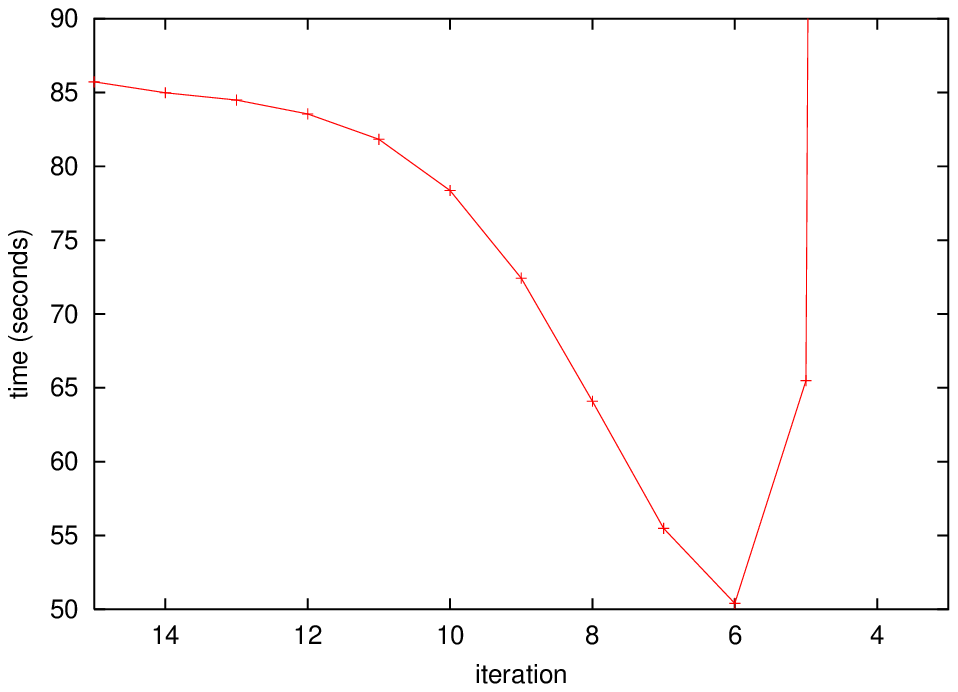}}
\caption{Combining iterations.}
\label{fig:kkcomb}
\end{figure}

We performed several experiments on each data set and limited the
number of candidate patterns that is allowed to be generated. If
the upper bound on the total number of candidate patterns is below
this limit, the algorithm generates and counts all possible
candidate patterns within the next iteration.
Figure~\ref{fig:kkcomb} shows the results. The $x$-axis shows the
total number of iterations in which the algorithm completed, and
the $y$-axis shows the total time the algorithm needed to
complete.
\addtocounter{figure}{-1}
\begin{figure}
\addtocounter{subfigure}{2}
\centering
\subfigure[T40I10D100K]{\includegraphics{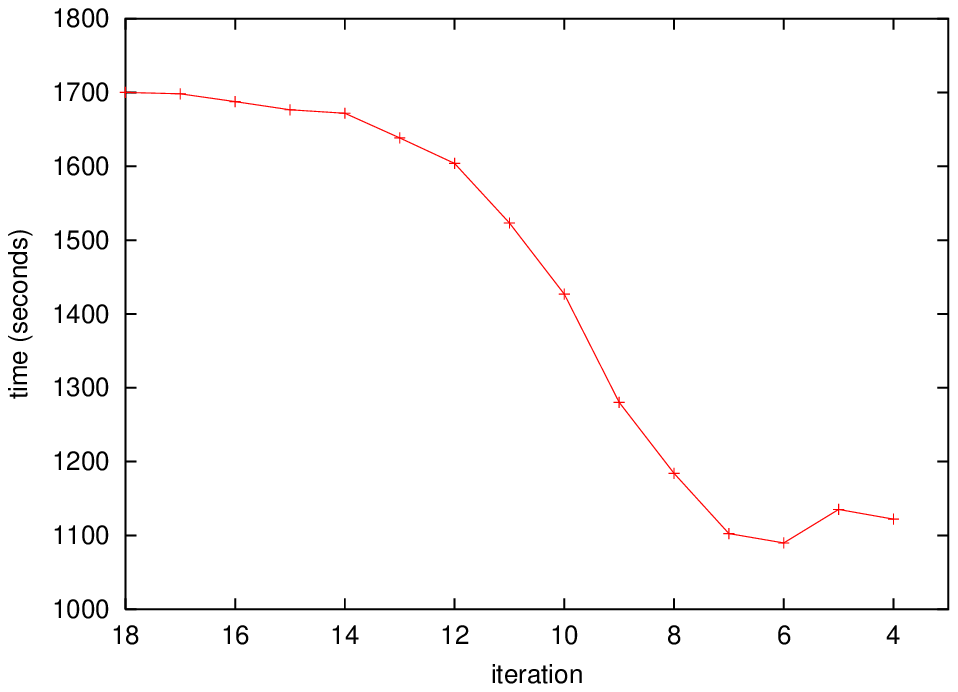}}\\
\subfigure[mushroom]{\includegraphics{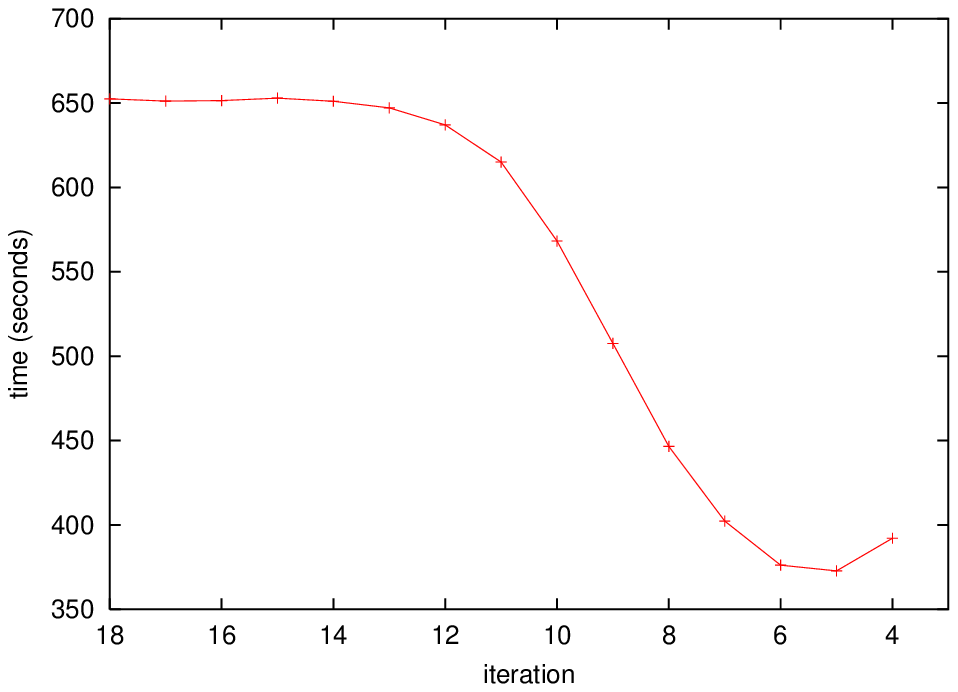}}
\caption{Combining iterations.}
\end{figure}
As can be seen, for all datasets, the algorithm can already
combine all remaining iterations into one very early in the
algorithm.  For example, the BMS-Webview-1 dataset, which normally
performs 15 iteration, could be reduced to six iterations to give
an optimal performance. If the algorithm already generated all
remaining candidate patterns in the fifth iteration, the number of
candidate patterns that turned out to be infrequent was too large,
such that the gain of reducing iterations has been consumed by the
time needed to count all these candidate patterns.  Nevertheless,
it is still more effective than not combining any passes at all.
If we allowed the generation of all candidate patterns to occur in
even earlier iterations, although the upper bound predicted a to
large number of candidate patterns, this number became indeed to
large keep in main memory.

\section{Conclusion} \label{conclusion}

Motivated by several heuristics to reduce the number of database
scans in the context of frequent pattern mining, we provide a hard
and tight combinatorial upper bound on the number of candidate
patterns and on the size of the largest possible candidate
pattern, given a set of frequent patterns. Our findings are not
restricted to a single algorithm, but can be applied to any
frequent pattern mining algorithm which is based on the levelwise
generation of candidate patterns. Using the standard Apriori
algorithm, on which most frequent pattern mining algorithms are
based, our experiments showed that these upper bounds can be used
to considerably reduce the number of database scans without taking
the risk of getting a combinatorial explosion of the number of
candidate patterns.

\section{Acknowledgement}

We wish to thank Blue Martini Software for contributing the KDD
Cup 2000 data, the machine learning repository librarians
Catherine Blake and Chris Mertz for providing access to the
mushroom data, and Tom Brijs for providing the Belgian retail
market basket data.

\bibliographystyle{plain}

\end{document}